\documentclass[aps,12pt,superscriptaddress,amsfonts,amssymb,amsmath,longbibliography]{revtex4-2}
\pdfoutput=1

\usepackage{graphicx}
\usepackage{xcolor}
\usepackage{amsmath}
\usepackage{bm}
\usepackage{amssymb}
\usepackage[a4paper,margin=2.6cm]{geometry}
\usepackage{hyperref}
\hypersetup{colorlinks=true,linkcolor=blue,citecolor=blue,urlcolor=blue}
\usepackage{float}

\begin{document}

\title{
Coherent Oscillations of Protons in Hydrogen-loaded Metals
}

\author{G.\ Modanese \footnote{Email address: giovanni.modanese@unibz.it}}
\affiliation{Free University of Bozen-Bolzano \\ Faculty of Engineering \\ I-39100 Bolzano, Italy}
\date{\today}

\linespread{0.9}

\begin{abstract}
We review recent calculations and numerical simulations showing the formation of coherent states of protons in hydrogen-loaded metals with a cubic crystal lattice \cite{GamberaleModanese2023Symmetry,GamberaleModanese2023PhysicaB,GamberaleModanese2024QuantumReports}. In these states protons oscillate coherently at frequencies of the order of $10^{13}\,\mathrm{Hz}$ and in a fixed phase relation with a strong high-frequency electric field which is trapped in the material, especially if the material is made of micro-powders.  
The energy gap of the coherent ground state is estimated to be well above
thermal energies, of the order of a fraction of an eV per particle, and therefore
large enough to make the state robust against thermal fluctuations.
The analytical calculations address the realistic case of a large number of protons, in the rotating-wave approximation. The numerical calculations are presently limited to a small number of protons but go beyond the rotating-wave approximation and allow one to take into account a dissipation term associated with the strong oscillating electric field. The next task of this theoretical model is to compute the excited states of the coherent system. There are strong indications that in those states protons can be excited by an external pump to energies much larger than those achievable by single incoherent protons in condensed matter. Such energies are large enough to make possible some electron-capture processes, with generation of slow neutrons. This dynamical mechanism offers an alternative to the Widom--Larsen hypothesis of ``heavy electrons'' \cite{WidomLarsen2006}, and is closer to current models in mainstream physics. The consequences, in terms of nuclear transmutations, of neutron generation via electron capture would be similar to those predicted by Widom and Larsen, plus several other processes like those recently re-examined by Metzler et al. \cite{Metzler2023}.
\end{abstract}

\maketitle

\section{Introduction}

A long-standing open problem in the physics of hydrogen-loaded metals is whether
collective quantum states of the absorbed hydrogen can play a direct role in
low-energy nuclear processes.  From a theoretical point of view, this question is
interesting independently of any specific phenomenological interpretation, because
hydrogen in a metallic lattice provides a dense system of light charged particles
bound to interstitial sites and coupled, in principle, to collective electromagnetic
modes of the host material.  If a sufficiently large number of such particles can
enter a coherent state, the relevant energy scales need not be limited to those of
single, incoherent particles.

This possibility is also motivated by a large body of experimental work.  Hundreds
of independent experiments have reported evidence for anomalous reactions occurring
in some hydride metals when subjected to thermal, electrical, or optical stimulation.
Nuclear transmutations of the metal atoms have been observed, and in some cases
also neutron emission and reactions of hydrogen isotopes leading, for example, to
the formation of tritium or other light elements 
\cite{Metzler2023,
Fomitchev2024,Iwamura2024,Hioki2024,Yamauchi2024,GamberaleModanese2024SymmetryDeuterium,
DeNinno1989,Celani2006}.  See also the tables of references in
Refs.~\cite{Metzler2023,Iwamura2024}.  Some of these reactions generate excess heat
of apparently non-chemical origin.  A satisfactory microscopic theory of such
phenomena is still lacking.  The aim of the present paper is not to provide a
complete reaction model, but to discuss one possible theoretical ingredient: the
formation and excitation of coherent proton states in hydrogen-loaded metals.

We shall focus on hydrogen-loaded systems, rather than deuterium-loaded systems.
This choice is deliberate.  From the point of view of many-body theory, deuterons
would be simpler, because they are bosons.  Protons, instead, are fermions, and one
must in principle worry about the exclusion principle.  However, the one-particle
wavefunctions of protons bound to tetrahedral or octahedral sites in a metallic
lattice are strongly localized, as shown in Ref.~\cite{GamberaleModanese2024QuantumReports}.
The overlap between wavefunctions centered at different sites is therefore very
small, so that an approximate boson-like treatment can be used for the collective
oscillatory degrees of freedom considered here.  This approximation would of course
be exact, as far as exchange symmetry is concerned, for deuterons.

There is also a physical reason for keeping the discussion centered on protons.
If a deuterium-loaded material contains collective states with sufficiently large
available energy, fusion channels are in principle also possible, although their
probabilities in a condensed-matter environment may be strongly suppressed.  In the
present work we want to isolate a different mechanism, namely electron capture by
coherent protons followed by neutron-induced nuclear processes.  In this mechanism
the elementary weak process is
\begin{equation}
        p+e \rightarrow n+\nu_e ,
\label{ECreaction}
\end{equation}
and the missing energy, approximately \(0.78\,\mathrm{MeV}\) for particles initially
at rest, is supplied by a collective excitation of the proton system.  The subsequent
neutron can then be captured by a light or heavy nucleus, producing deuterium,
tritium, or a transmuted isotope, depending on the material and on the nuclear
channel involved.  Thus the role of the coherent proton state is not to enhance
ordinary fusion, but to provide, at least in principle, a collective energy reservoir
for electron capture.

The formation of the coherent ground state has been studied in
Refs.~\cite{GamberaleModanese2023Symmetry,GamberaleModanese2023PhysicaB}.  In that model,
a large number \(N\) of protons are bound to interstitial sites of a cubic lattice
and oscillate with a local frequency determined by the electrostatic potential of
the site.  The protons are also coupled to electromagnetic modes confined inside a
finite region of the material.  Above a density-dependent threshold, the system can
lower its energy by forming a collective state in which the proton oscillations are
phase-correlated with a self-generated electromagnetic field.  The coherent ground
state is therefore separated from the incoherent perturbative ground state by an
energy gap, which makes it robust against thermal fluctuations.

For electron capture, however, the ground-state gap alone is not sufficient.  One
needs excited coherent states capable of storing and transferring an energy of order
\(0.78\,\mathrm{MeV}\) to a single electron-capture event.  The basic idea is that
collective excitation energies may scale with the number of participating protons,
and can therefore be much larger than the level spacings of individual oscillators.
This is qualitatively illustrated in Figs.~\ref{fig:phase1}--\ref{fig:phase4_5}.
A related phenomenon, in which a large quantum is redistributed among many smaller
degrees of freedom, is often referred to as down-conversion or fractionation of a
large quantum, and has been studied theoretically in other condensed-matter contexts,
for example by Hagelstein and Chaudhary~\cite{HagelsteinChaudhary2008Level}.  More
recently, Ref.~\cite{GamberaleModanese2025Annalen} has provided numerical evidence,
in a simplified model of coherent harmonic qubits, that high-frequency energy
transfer between external oscillators can occur only when the intermediate
low-frequency oscillators are in a coherent state.  Although this simplified model
does not yet constitute a full calculation for hydrogen-loaded metals, it supports
the central assumption that coherent excited states can possess transition energies
much larger than those available to isolated incoherent oscillators.

The large energy gaps of the collective proton system could also play a role after
electron capture has occurred.  The slow neutrons produced by EC can be absorbed by
heavy nuclei of the host metal, causing transmutations, or by light nuclei.  For
hydrogen loading, an important channel is
\begin{equation}
        p+n\rightarrow D+\gamma ,
\end{equation}
with emission of a \(2.2\,\mathrm{MeV}\) gamma ray.  For deuterium loading, neutron
capture can lead to tritium formation.  In both cases, the gamma energy may be
partially reabsorbed by collective degrees of freedom of the proton or deuteron
system, although the efficiency and kinetics of such processes are presently
unknown.  These issues require knowledge of the spectrum and transition probabilities
of the excited coherent states, which remains one of the main open problems of the
model.

The occurrence of EC in condensed matter has been previously suggested by Widom and Larsen \cite{WidomLarsen2006}. In the weak interaction process \eqref{ECreaction}
the missing energy in case of initial particles at rest was hypothesized to come from the increase of the effective electron mass due to the electromagnetic electron interaction in matter. Proton oscillations were supposed to play only an indirect role, because they generated strong oscillating fields on the metal surface and these in turn increased the effective electron mass. This hypothesis of the ``heavy electron'' is hard to prove, for several reasons, and has been strongly criticised \cite{HagelsteinChaudhary2008Mass,Hagelstein2013,Ciuchi2012,Pourjafarabadi2022}. Widom and Larsen defended their proposal also in other articles \cite{WidomLarsen2006Nuclear,Widom2008,Srivastava2012} and described a possible suppression mechanism of the gamma rays which originate from transmutations of heavy elements and reactions such as $p+n\to D+\gamma$ and $D+n\to T+\gamma$. This suppression mechanism is also based on the heavy-electron hypothesis. Note that the generation of tritium has been observed in several experiments, mainly starting from deuterium but also from hydrogen \cite{Storms1990,Will1993,Claytor1998,Szpak1998,Packham1989,Bockris1992,Chien1992,Srinivasan2015}.

The electron capture by coherent protons as proposed in this paper has been
previously analyzed by Gamberale~\cite{Gamberale2022}.  The purpose of that
work was complementary to the one pursued in Refs.~\cite{GamberaleModanese2023Symmetry,
GamberaleModanese2023PhysicaB}.  Instead of deriving the coherent state from the
microscopic matter--field Hamiltonian, Gamberale assumed the existence of coherent
proton and electron plasmas and used them as initial and final states in a
quantum-field-theoretical calculation of the weak process \eqref{ECreaction}.
This made it possible to address a question which is not treated in the present
review of the coherence mechanism itself, namely the order of magnitude of the
neutron production rate once a coherent excitation above the electron-capture
threshold is available. The work also includes rate equations relating the pumping process to the final energy output, which are clearly relevant to applications.

A central point of Ref.~\cite{Gamberale2022} is that the missing energy for
electron capture need not be supplied by a single particle.  If a coherence
domain contains a very large number \(N\) of charged particles, a collective
variation of the energy per particle of order
\(
        \Delta E/N
\)
can correspond to a macroscopic energy \(\Delta E\) available for one weak
transition.  In this sense the threshold
\(
        m_n-m_p-m_e\simeq 0.782\,{\rm MeV}
\)
is not crossed by accelerating one proton or one electron to MeV energies, but
by a small de-excitation of a many-body coherent state. This is conceptually close to the
``fractionation'' or down-conversion idea discussed above, but it is implemented
there at the level of the weak-interaction matrix element.

The treatment of the excited coherent state in Ref.~\cite{Gamberale2022} is based
on an Ansatz inspired by Preparata's coherent-electrodynamics approach
\cite{Preparata1995}.  The coherent ground state is parametrized by a collective
mixing angle, fixed by minimization of the energy.  A coherent excited state is
then represented by a nearby state in the same class, with slightly different
values of the collective parameters, for example a modified mixing angle and a
modified collective momentum.  Thus the excitation is not described as the
promotion of one proton to a high single-particle level, but as a small collective
displacement of the entire coherent configuration.  The electron-capture final
state is obtained by applying the appropriate destruction operators to this
many-body coherent state, so that one proton and one electron are removed from
the coherent domain while the large-\(N\) coherent structure is otherwise
preserved.

This Ansatz should be regarded as physically motivated but not yet as a derived
spectrum of the microscopic proton--photon Hamiltonian.  This is where
the later numerical work on coherent harmonic qubits~\cite{GamberaleModanese2025Annalen}
may become relevant.  In Ref.~\cite{GamberaleModanese2025Annalen} the excited
states are not postulated through a Preparata-type variational form.  Instead,
a simplified system of \(N\) two-state oscillators coupled to a dissipative photon
mode is probed dynamically by weakly coupled external oscillators of variable
frequency.  Resonant absorption and resonant transfer between two external
oscillators then provide operational information on the transition energies and
transition probabilities of the coherent system.  The important qualitative result
is that high-frequency transfer mediated by low-frequency oscillators appears only
in the coherent phase.  This supports the basic physical assumption used in
Ref.~\cite{Gamberale2022}, namely that coherent excited states can exchange energy
collectively at frequencies much higher than those of the individual oscillators,
but it also shows that a quantitative theory of such states should ultimately be
based on their actual spectrum and transition matrix elements rather than on an
Ansatz alone.

The comparison also clarifies the present status of the theory.  The calculation
of Ref.~\cite{Gamberale2022} is the most direct attempt to connect coherent
proton states with the weak electron-capture rate, but it relies on an assumed
structure of the excited coherent states.  The numerical results of
Ref.~\cite{GamberaleModanese2025Annalen} point in the same physical direction,
showing collective high-frequency transitions in a simplified model, but they do
not yet provide the full excited-state spectrum of a realistic hydrogen-loaded
metal.  A complete theory of coherent electron capture should therefore combine
the microscopic derivation of the coherent ground state, the localized proton
spectrum of Ref.~\cite{GamberaleModanese2024QuantumReports}, and a many-body calculation
of the excited coherent states and of their weak-interaction matrix elements.

\section{Elements from the calculation ``Coherent plasma in a lattice''}

In Refs.~\cite{GamberaleModanese2023Symmetry,GamberaleModanese2023PhysicaB}, we considered a system with a large number $N$ of protons bound in potential wells centered at tetrahedral sites $x_1,\ldots,x_N$ in a cubic metal lattice. All protons oscillate with the same elastic frequency $\omega$ defined by the local electrostatic potential; in addition, they are subject to collective plasma oscillations with a frequency $\omega_p$ which depends on the proton density and is of the same magnitude as $\omega$. The protons also interact with cavity photons, i.e. electromagnetic modes confined in a finite volume of size $\sim\lambda^3$, where $\lambda$ is the resonant photon wavelength, related to $\omega$ and $\omega_p$.

\begin{figure}[t]
    \centering
    \includegraphics[width=0.75\linewidth]{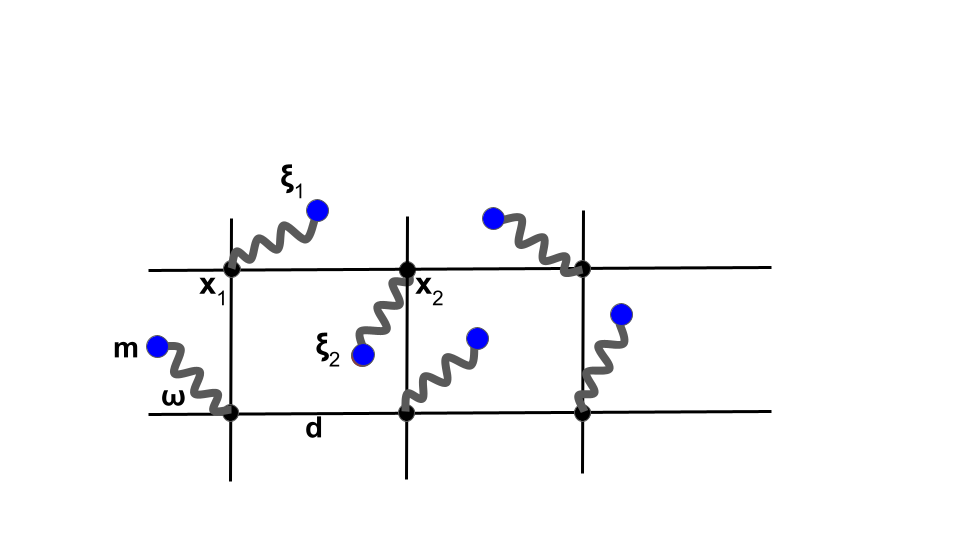}
    \caption{Notations for particles of mass $m$ bound at the lattice sites $x_1,x_2,\ldots,x_N$ by a local elastic potential with oscillation frequency $\omega$.}
    \label{fig:lattice_notation}
\end{figure}

The one-particle energy spectrum of protons bound by the local electrostatic potential has been computed numerically in Ref.~\cite{GamberaleModanese2024QuantumReports} for a typical lattice with unit spacing $d=0.25\,\mathrm{nm}$ and a frequency $\hbar\omega=0.41\,\mathrm{eV}$. This corresponds to $\lambda\sim 3\,\mu\mathrm{m}$. Note that a cube of side $1\,\mu\mathrm{m}$ contains $64\times 10^{9}$ cells. It turns out to have 12 bound states with equal gaps in the energy interval $[-4.8\,\mathrm{eV},0]$, and a continuum of unbound states above zero. The bound eigenfunctions are strongly localized in the tetrahedral site wells, due to the large mass of protons. This justifies an approximated boson-like treatment without Pauli exclusion. The calculation can of course be repeated for deuterons instead of protons, in which case the boson-like treatment is exact.

\begin{figure}[t]
    \centering
    \includegraphics[width=0.78\linewidth]{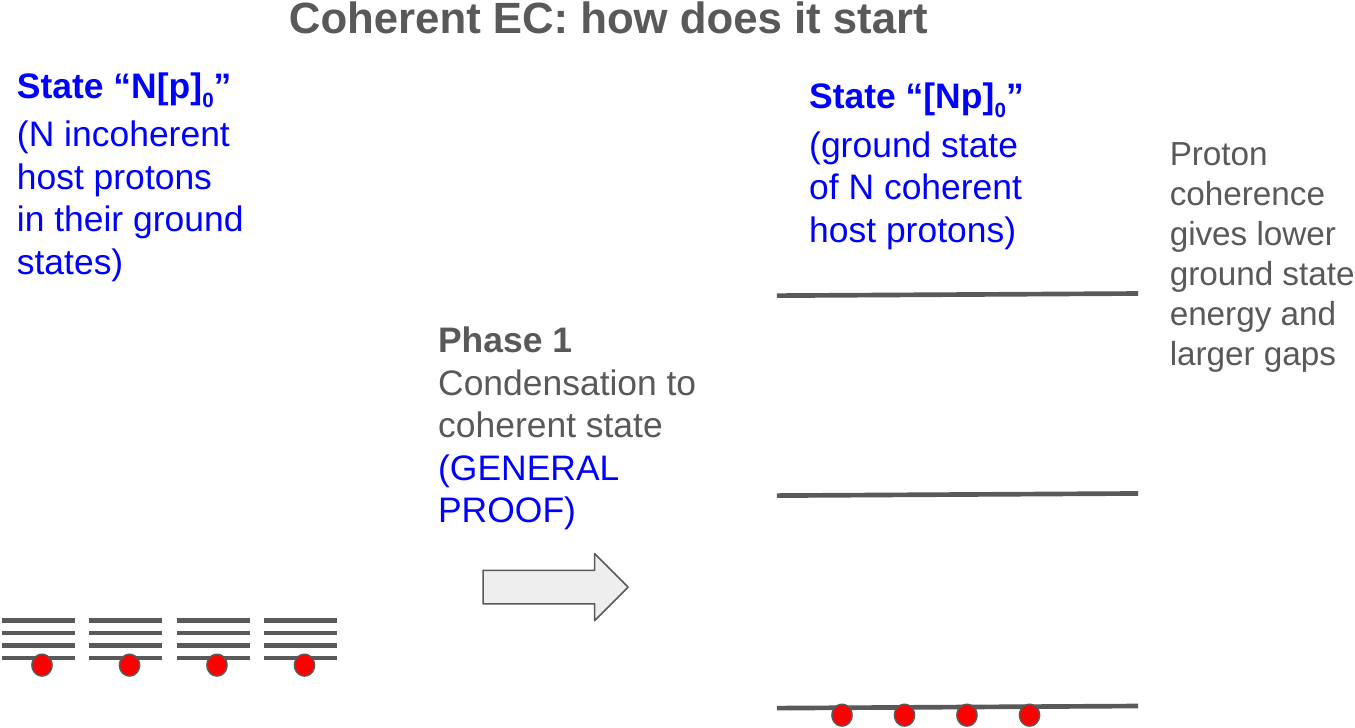}
    \caption{Graphical representation of the coherent electron-capture process investigated in Refs.~\cite{GamberaleModanese2023Symmetry,GamberaleModanese2023PhysicaB,GamberaleModanese2024QuantumReports}. A process of this kind might be responsible for generation of slow neutrons in a metal hydride, for example TiH$_x$, with subsequent production of deuterium and isotopes of Ti. Phase 1: following the absorption of hydrogen in Ti, protons are bound to tetrahedral or octahedral sites in the Ti crystal structure, with an approximately harmonic potential. The wavefunctions of the single protons become synchronized, through a complex mechanism analyzed in Refs.~\cite{GamberaleModanese2023Symmetry,GamberaleModanese2023PhysicaB,GamberaleModanese2024QuantumReports}, and protons condense into a collective coherent state with an energy gap well above thermal energies, robust with respect to thermal fluctuations. The number $N$ of protons in a single coherence domain can be quite large, up to $N\sim 10^{11}$. The coherent ground state of $N$ protons is denoted here by $[Np]_0$. The coherent excited states have energies much larger than the energy levels of the single oscillators. Note that in the picture only a few protons and a few energy levels have been represented.}
    \label{fig:phase1}
\end{figure}

\begin{figure}[t]
    \centering
    \includegraphics[width=0.78\linewidth]{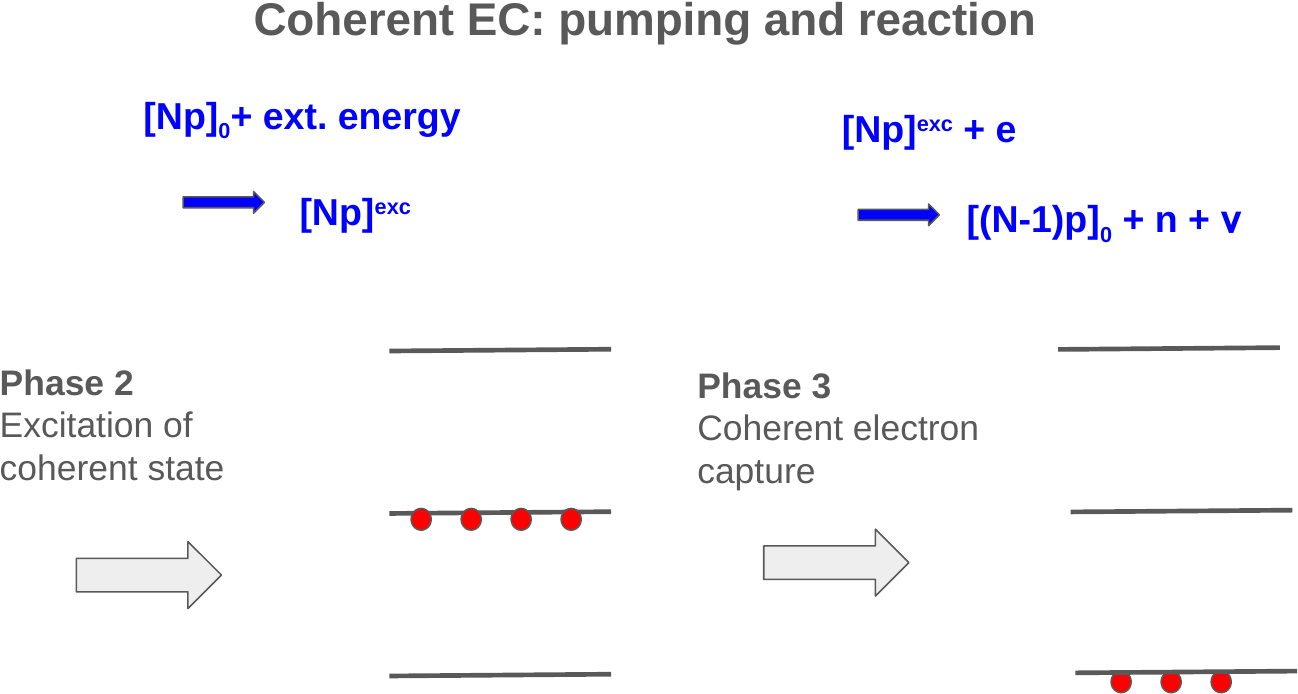}
    \caption{Phase 2: when power is supplied by a suitable external source, the system evolves into a collective excited state. The energy of this state is higher than the threshold for production of a slow neutron in an electron-capture process, Phase 3, in which the number of protons decreases by one, a neutrino is generated, and the collective proton state is de-excited.}
    \label{fig:phase2_3}
\end{figure}

\begin{figure}[t]
    \centering
    \includegraphics[width=0.78\linewidth]{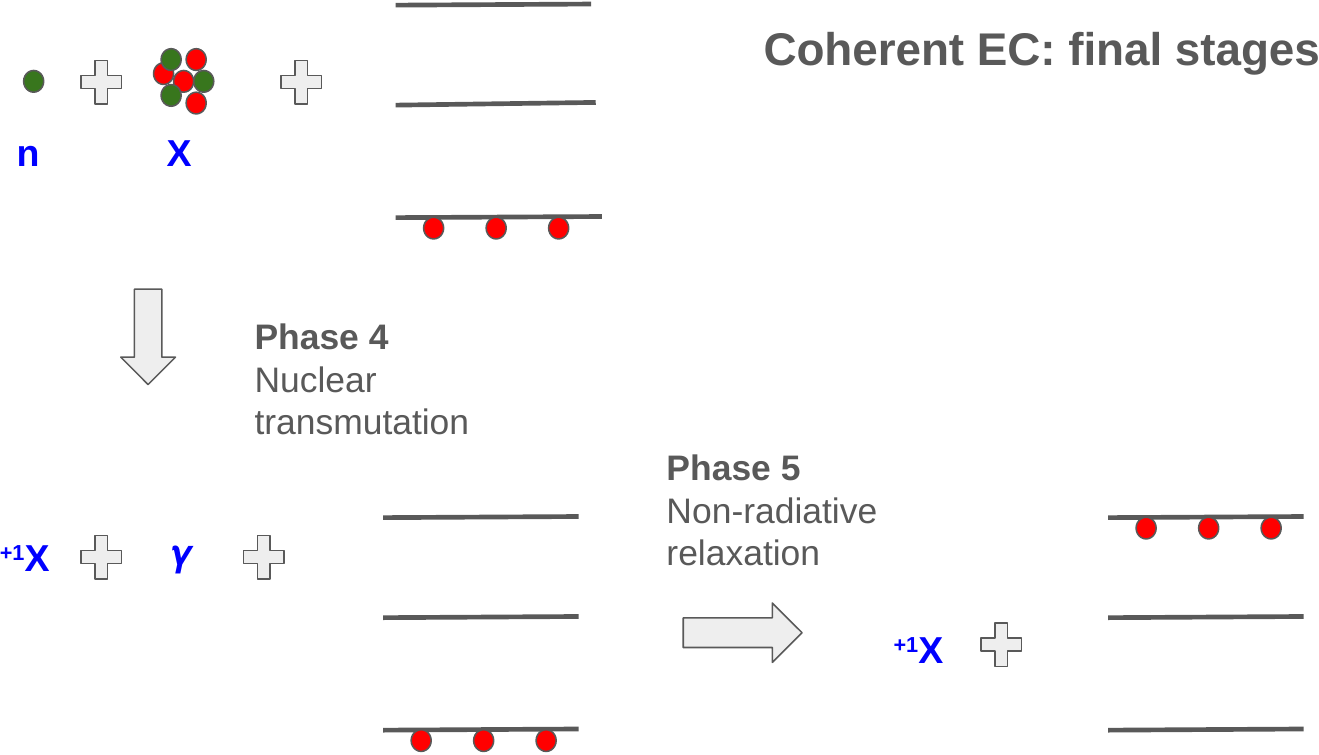}
    \caption{Phase 4: the neutron causes a transmutation in which the mass of an absorbing nucleus $X$, in our case Ti or H, increases by one and energy is released in a gamma ray. Phase 5: the gamma ray may be partially absorbed or reprocessed by collective degrees of freedom. In favorable conditions, the excited coherent state could support further electron captures, even if the external power supply is not available anymore.}
    \label{fig:phase4_5}
\end{figure}

Low-energy nuclear reactions via coherent EC can be schematically thought to occur in five stages, depicted in Figs.~\ref{fig:phase1}--\ref{fig:phase4_5}: Phase 1, condensation to coherent state; Phase 2, excitation of the coherent state; Phase 3, coherent EC; Phase 4, nuclear transmutation; Phase 5, non-radiative relaxation. The theoretical model with analytical and numerical calculations of Refs.~\cite{GamberaleModanese2023Symmetry,GamberaleModanese2023PhysicaB,GamberaleModanese2024QuantumReports} offers a rigorous proof of Phase 1.

Introducing some more formal details, the Hamiltonian of the bound protons is
\begin{equation}
    H_{\mathrm{osc}}=\omega \sum_{n=1}^{N_{\mathrm{osc}}}
    \left(a_n^\dagger a_n+\frac{3}{2}\right),
    \label{eq:Hosc}
\end{equation}
where
\begin{equation}
    \omega' = \sqrt{\omega^2+\omega_p^2}
    \label{eq:omega_ren}
\end{equation}
is the frequency renormalized by plasma oscillations, and all operators have an implicit time dependence in the interaction representation.

The Hamiltonian of the photons in a finite volume $V$ requires a careful treatment of photon modes and their frequencies, according to general quantum-optics rules. Here we can just sketch a few essential points. A priori, photon modes with frequencies $\omega$, $\omega_p$, or $\omega'$ are involved in resonant interactions. One starts by writing the pure photon part of the modes of frequency $\omega$:
\begin{equation}
    H_{\mathrm{photon}} = \omega \sum_{p,\mathbf{k}}
    \left(b_{p,\mathbf{k}}^\dagger b_{p,\mathbf{k}}+\frac{1}{2}\right),
    \qquad |\mathbf{k}|=\omega .
    \label{eq:Hphoton}
\end{equation}
Here the summation is taken over two photon polarizations $p$ and all possible directions of $\mathbf{k}$, i.e. over solid angles.

The matter-field interaction is written as a dipole interaction plus diamagnetic terms by replacing the momentum $\mathbf{p}$ by $\mathbf{p}+e\mathbf{A}$ in the Hamiltonian. One obtains
\begin{align}
    H_{\mathrm{dip}} &= \sum_{n=1}^{N_{\mathrm{osc}}}\frac{e}{m}\,
    \mathbf{p}_n\cdot\mathbf{A}(\mathbf{x}_n+\bm{\xi}_n,t),
    \label{eq:Hdip}\\
    H_{A^2} &= \sum_{n=1}^{N_{\mathrm{osc}}}\frac{e}{2m}\,
    \mathbf{A}^2(\mathbf{x}_n+\bm{\xi}_n,t).
    \label{eq:HA2}
\end{align}
Next, two crucial mathematical transformations of the photon field are performed, preserving the commutation rules. The first is a linear combination known to eliminate the diamagnetic term,
\begin{equation}
    c_{p,\mathbf{k}} = \frac{1}{2\sqrt{\omega\omega'}}
    \left[(\omega'+\omega)b_{p,\mathbf{k}}+(\omega'-\omega)b_{p,\mathbf{k}}^\dagger\right].
    \label{eq:c_pk}
\end{equation}
This clearly makes sense only if $\omega$ is not zero. In our case, $\omega$ is defined by the electrostatic lattice forces, while often in quantum optics it is defined by the size of the external artificial cavity. The second step is the summation of modes in all space directions,
\begin{equation}
    C_i = \sqrt{\frac{3}{8\pi}}\sum_{p,\mathbf{k}} (\hat{\mathbf{e}}_i\cdot\mathbf{e}_{p,\mathbf{k}})c_{p,\mathbf{k}} .
    \label{eq:Ci}
\end{equation}
This introduces a three-dimensional improvement factor $\sqrt{8\pi/3}$ which is crucial for condensation.

The formation of resonant electromagnetic modes in all three space directions is possible because the finite volume acts as a natural cavity. Inside one micro-grain of the metal, the bare photon frequency $\omega=c k$ changes to $\omega'>\omega$ by combination with the plasma frequency, while the photon momentum $\mathbf{k}$ is not affected. An external photon mode has a different ratio of energy and momentum. This mismatch prevents internal modes from propagating outside the material and thus provides a justification of the three-dimensional cavity effect. Possible losses at the interface can be taken into account numerically for small $N$, in addition to ohmic losses in the bulk. One finds that the occurrence of condensation is not affected by the losses.

The total effective Hamiltonian can be rewritten as
\begin{align}
    H_{\mathrm{tot}} &= H_{\mathrm{osc}} + \omega' \sum_{i=1}^{3}
    \left(C_i^\dagger C_i+\frac{1}{2}\right) \nonumber\\
    &\quad + \frac{i\omega_p}{2\sqrt{N_{\mathrm{osc}}}}\sqrt{\frac{8\pi}{3}}
    \sum_{n=1}^{N_{\mathrm{osc}}}\sum_{i=1}^{3}
    \left[a_{n,i}^\dagger C_i-a_{n,i}C_i^\dagger+a_{n,i}^\dagger C_i^\dagger-a_{n,i}C_i\right].
    \label{eq:Htot}
\end{align}
After minimization, the energy gap per particle of the coherent state is found to be
\begin{equation}
    \delta E_{\Omega}^{(1)} = \omega' |\alpha|^2
    \left(1-\frac{2\pi}{3}\epsilon^2\right),
    \label{eq:gap}
\end{equation}
where $\epsilon$ is a dimensionless coupling and $\alpha$ is the maximum oscillation amplitude of the protons, defined by the number of bound eigenstates in the electrostatic potential wells.

The work in Ref.~\cite{GamberaleModanese2025Annalen} concerns the numerical computation of the excited states of coherent oscillators in the simple case of quantum oscillators with two states only, called ``harmonic qubits''. Information on the spectrum of these states and on their transition probabilities is obtained by simulating their coupling to one or two external oscillators with higher frequency. The model comprises $N$ qubits coupled to a self-generated photon field that is able to synchronize them into a coherent ground state, even in the presence of dissipation. The qubits are additionally coupled to a variable external high-frequency oscillator $Q$. In correspondence with the frequency of a collective excited state of the qubits, energy is absorbed from the external oscillator. If two variable external oscillators $Q_1$ and $Q_2$ are coupled to each other via the coherent qubits, energy transfer between them is possible when their frequency matches the frequency of an excited coherent state. The published calculations show that high-frequency energy transfer between two external oscillators mediated by a system of low-frequency oscillators is only possible when this system is in a coherent state.

\section{Conclusions and outlook}

The formation of coherent proton states in highly hydrogenated metals is a plausible theoretical mechanism for creating collective energy reservoirs in condensed matter.  If suitable excited coherent states exist, they could provide the energy required for electron capture, with generation of slow neutrons.  These neutrons could then be captured by light or heavy nuclei, leading to deuterium or tritium formation, isotope shifts, or nuclear transmutations.  Gamma rays emitted in such secondary reactions might be partially reabsorbed or reprocessed by collective degrees of freedom of the coherent proton system, but the corresponding efficiencies and reaction kinetics are presently unknown.

The central open problem is the structure of the excited coherent states.  The existence of the coherent ground state has been derived analytically, and the localization of the single-proton bound states has been studied numerically.  However, electron capture requires not only a stable coherent ground state, but also excited coherent states with suitable transition energies and transition strengths.  In the calculation of Ref.~\cite{Gamberale2022}, the excited states were described by a Preparata-type Ansatz.  This provides a useful first estimate of the electron-capture rate, but a more complete theory should derive the excited spectrum from the microscopic proton--photon Hamiltonian.

The size of a coherence domain is expected to be at most of the order of the wavelength of the resonant electromagnetic mode trapped in it, namely $\sim 1\,\mu\mathrm{m}$.  This corresponds to $\sim 10^{10}$--$10^{11}$ lattice sites.  In principle, such domains may form spontaneously, as proposed in coherent electrodynamics models of water \cite{DeNinnoGamberale2025,Bono2012}, although an analogous proof for hydrogen-loaded metals is still missing.  In practice, a granular or microstructured material could help define effective cavities, possibly even on scales smaller than the wavelength, since the minimum number of coherent protons needed to supply the energy for one electron-capture event is much smaller, of order $10^6$.

It is not clear whether coherence domains could be observed directly through their interaction with external gamma rays.  In principle, a coherent domain with a suitable spectrum of collective excited states could have an enhanced probability of absorbing or reprocessing gamma radiation, especially if the gamma energy can be redistributed among many low-energy degrees of freedom.  However, the experimentally measured attenuation or re-emission signal would be an average over the whole sample.  If, at a given time, only a small fraction of the material volume is occupied by active coherence domains, the corresponding signal would be strongly diluted and could be hidden by ordinary photoelectric, Compton, and nuclear absorption processes in the metal.  Moreover, the domains may be transient and spatially inhomogeneous, so that a null result in a bulk gamma-absorption experiment would not necessarily rule out their existence.  Conversely, a positive result would require careful separation from standard gamma-interaction mechanisms.

Preliminary simulations with a small number of particles suggest the existence of a discrete spectrum in which some transition energies grow with the number of particles participating in the coherent state.  In order to allow electron capture, the coherent system must first be driven into an excited state.  The external pump, however, is expected to act at frequencies of the order of the base oscillation frequency $\omega$, rather than directly at the much larger collective transition frequency.  The relevant physical problem is therefore a form of coherent energy conversion between strongly mismatched energy scales.

Related issues appear in the spin--boson models studied by Hagelstein and Chaudhary \cite{HagelsteinChaudhary2008Level,HagelsteinChaudhary2008SpinOne}, where a large two-level transition can become resonant with many quanta of a low-frequency oscillator through multiphoton Bloch--Siegert anticrossings.  Those models show that an energy match alone is not sufficient: one must also compute the level splitting, or equivalently the transition matrix element, because this determines whether energy exchange can occur on a physically relevant time scale.  The analogy with the present problem is limited, since the spin--boson models involve few degrees of freedom and a prescribed oscillator mode, while the coherent-proton model involves many localized proton oscillators, a self-generated electromagnetic field, and dissipation.  Still, they provide useful guidance for future work: coherent down-conversion requires a precise resonance condition, a non-negligible anticrossing gap, and sufficient stability of the resonance during the exchange of many low-energy quanta.

Experiments on ensembles of $^{57}\mathrm{Fe}$ nuclei excited by x-ray free-electron-laser pulses \cite{Chumakov2018} provide a different but relevant benchmark.  They demonstrate collective nuclear response and photon-by-photon superradiant decay in an extended solid-state system.  They do not demonstrate the down-conversion mechanism required here, but they show that coherent nuclear ensembles in crystals can display measurable collective radiative dynamics.

The comparison between Ref.~\cite{Gamberale2022} and the later work on coherent harmonic qubits \cite{GamberaleModanese2025Annalen} clarifies the present status of the theory.  Ref.~\cite{Gamberale2022} gives the first explicit weak-interaction calculation of neutron production from coherent proton states, but it relies on an assumed form of the excited coherent configurations.  The numerical results of Ref.~\cite{GamberaleModanese2025Annalen}, although obtained in a simplified two-level model, point in the same physical direction by showing that high-frequency energy transfer can be mediated by a coherent system of low-frequency oscillators.  A complete theory should combine these two ingredients: the weak-interaction matrix element of Ref.~\cite{Gamberale2022} with a microscopic calculation of the excited coherent spectrum and of the corresponding transition strengths.

In modern many-body language, the ultimate quantity to be computed is not only the total energy stored in a coherence domain, but the spectral weight with which the local weak operator couples the initial coherent state to final states containing a neutron and a neutrino.  This requires a dynamical response function or structure factor of the coherent medium.  Establishing whether such a response contains a coherent contribution scaling super-extensively with the number of participating protons remains an open problem.

Finally, weak-interaction mechanisms for low-energy nuclear processes have also been considered in a different framework in Ref.~\cite{Ramkumar2024}, where proton-to-neutron conversion is treated as the first step of a second-order perturbative process.  In that approach the neutron is a virtual intermediate state subsequently captured by a nucleus, and no coherent proton domain is assumed.  The rate is found to be negligible in general, but it may be enhanced in the presence of a suitable nuclear resonance.  This is conceptually different from the coherent-proton mechanism discussed here, where the missing energy is supplied by a real collective excitation of the medium.  Nevertheless, the comparison is instructive: in both cases an observable rate requires a non-trivial spectral enhancement, either a nuclear resonance or a coherent many-body transition with sufficient spectral weight.

\bibliographystyle{apsrev4-2}
\bibliography{collective_oscillations_protons}

\end{document}